\documentclass[prd,onecolumn,nofootinbib]{revtex4}
\usepackage{bm,amsmath,amssymb,graphicx}
\begin{document}

\title{Parametric equations of the expansion of a closed universe with torsion}
\author{Prastik Mohanraj}
\affiliation{Engineering and Science University Magnet School, 500 Boston Post Road, West Haven, Connecticut 06516, USA}
\email{prastikmohanraj@gmail.com}

\begin{abstract}
\noindent
Using the Friedmann equations for a closed universe with spin and torsion, parametric equations for the scale factor and time are derived for the early universe.
\end{abstract}

\maketitle

\section{Introduction}
\noindent
Analogous to the Doppler effect perceived with sound waves, where approaching sound waves increase in pitch and receding sound waves decrease in pitch, the Doppler effect can be perceived with electromagnetic radiation, or light, where approaching waves increase in frequency and receding waves decrease in frequency. An increase in frequency corresponds to "bluer" light, making approaching waves experience blueshift. Similarly, receding waves decrease in frequency, producing "redder" light, and thus experiencing redshift. 
\newline
\newline
When galaxies are observed in the sky, over time the light coming from them becomes redder in color, thus these galaxies are experiencing redshift \cite{Ric}. In fact, all galaxies in the universe except the galaxies in the neighborhood of our own Milky Way appear to experience redshift. As a result, it can be concluded that every galaxy is receding from our perspective. The recession of galaxies is described by Hubble's law, which dictates that the speed of recession of a point in space, relative to any observer, is directly proportional to the distance between the two points. This law, together with the assumption that the universe is homogeneous and isotropic at largest scales (the cosmological principle), leads to a conclusion that the space itself is stretching by a constant factor over time, thus increasing distance between two points and causing proportional recession between these points. It can be pictured as how two points on a balloon seem to be moving away from each other when the balloon expands, although the points themselves are not experiencing motion, but the surface area on the balloon itself is increasing.
\newline
\newline
Since all these points are receding from one another, in that the space between these points is rapidly increasing, the universe is expanding. If the universe is expanding, it must have been much smaller, hotter, and denser in the distant past. Thus, it is likely that the universe started from an infinitesimally small size, and later undergone rapid expansion to create the universe's size we observe today. This idea is the foundation of the standard Big Bang cosmology.
\newline
\newline
There are currently three models of a homogeneous and isotropic universe: positively curved universe, negatively curved universe, and universe with zero curvature \cite{Lor}. The 3D plane of the universe becomes curved around the 4D hyperspace of the universe in one of these three models, much like the 2D surface of a balloon wraps around the 3D interior space of the balloon. A negatively curved universe is open, saddle shaped, with a lower energy density than the critical energy density. A universe with zero curvature is a flat, Euclidean space with its energy density equal to the critical energy density. A positively curved universe is closed and spherical, with higher energy density than the critical energy density. In this paper, the positively curved, hyper-spherical universe model will be analyzed.
\newline
\newline
For a closed universe, the scale factor $a(t)$ is is a function of the cosmic time $t$, acting as the radius of the 4D hyper-sphere whose 3D hyper-surface is the closed universe itself. Hubble's law states that the rate of expansion of the universe obeys $\dot{a}=H\cdot a$, where $H$ is the Hubble parameter and a dot denotes the derivative with respect to $t$. The critical energy density is equal to $\epsilon_c=\frac{(3H^2)}{(8\pi G)}$. Units will be used such that the speed of light $c=1$.
\newline
\newline
This paper establishes the novel finding of a closed universe with spin and torsion in the Einstein-Cartan theory of gravity, which extends general relativity to include spin of elementary particles. At extremely high densities, torsion manifests itself as a gravitational repulsive force, which avoids the Big-Bang singularity and replaces the Big Bang with a bounce that never reaches a singularity. We found the solution to the equation which describes the expansion of the universe right after the bounce, in the torsion-dominated era.
\newline
\newline
In Section II, we will see the Friedmann equations and their forms in the different stages of the universe's evolution. In Section III, we will examine the matter-dominated universe stage of evolution, and present the cycloid as a solution to this evolutionary stage. In Section IV, we will edit the cycloid to form the trochoid, and present the novel solution of using a trochoid for the torsion-dominated era of the universe. In Section V, we will apply the torsion-dominated trochoid model to the radiation-dominated model, where the torsion factor has cancelled out, and finally, in Section VI, we will explore the last stage of evolution, the dark-energy-dominated universe, with the Friedmann equations. Section VII provides a conclusion and summary of the analyses presented here.
\newline
\newline
We will now introduce the Friedmann equations, a background of cycloids, which will be crucial to the following analyses, and solve for parametric equations in the different eras of the universe, ultimately reaching the parametric equations for the beginning of the universe during the torsion-dominated era.
\section{Friedmann Equations}
\noindent
The Friedmann equations are an extension of the general theory of relativity, applied to the universe on a cosmological scale. These equations can be analyzed to develop equations of scale factor as a function of time. The easiest approach is to develop parametric equations of scale factor and time, in terms of another variable. For reasons later described, these parametric equations will be derived as a result of similarities between the Friedmann equations for a positively curved universe and the derivation of parametric equations for cycloidal/trochoidal motion.
\newline
\newline
The first Friedmann equation can be written as \cite{LL,Lor}
\begin{equation}
(\dot{a})^2+k=\frac{1}{3}  \kappa\epsilon a^2.
\label{original}
\end{equation}
Here, $\epsilon$ is the energy density of matter in the universe, dominated by relativistic light energy in the early universe and non-relativistic energy of massive particles in the later universe. Also, $\kappa=8\pi G$. In this analysis, the universe will be defined as having positive curvature, thus $k=+1$. The second Friedmann equation can be written as \cite{ApJ}
\begin{equation}
(\dot{\epsilon a^3})+p(\dot{a^3})=f(a,\epsilon),
\label{secondoriginal}
\end{equation}
where $p$ is the pressure of matter. The term $f(a,\epsilon)$ is effectively equal to $0$, or is adiabatic, for all stages of the universe except for the earliest stages where quantum production of particles increases entropy.
\newline
\newline
For relativistic matter, $p=\frac{1}{3}\epsilon$, so Equation (\ref{secondoriginal}) can be modified to yield $a \dot{\epsilon}+4\dot{a} \epsilon=0$, thus $\epsilon\propto \frac{1}{a^4}$, thus
\begin{equation}
(\dot{a})^2+1=\frac{2A}{a^2}
\label{radiationdominated}
\end{equation}
for the radiation-dominated era of the universe, where the positive constant $A\propto \frac{1}{3}\kappa$, or $A\propto \frac{8}{9}\pi G$. Similarly, for non-relativistic matter, $p=0$, so Equation (\ref{secondoriginal}) can be modified to yield $a\dot{\epsilon}+3\dot{a}\epsilon=0$, thus $\epsilon\propto \frac{1}{a^3}$, thus
\begin{equation}
(\dot{a})^2+1=\frac{2C}{a}
\label{matterdominated}
\end{equation}
for matter-dominated era of the universe, where the positive constant $C\propto \frac{1}{3}\kappa$, or $C\propto \frac{8}{9}\pi G$. The usage of coefficients $2A$ and $2C$ as opposed to $A$ and $C$ is simply for convenience.
\newline
\newline
These equations, both with the term $(\dot{a})^2+1$, resemble the parametric equations of $x$ and $y$ coordinates of a cycloid/trochoid graphed on a Cartesian plane. As a result, scale factor $a$ and time $t$ will be solved for in terms of parametric equations with a common variable $\theta$, similar to parametric equations of cycloids/trochoids with variable $\theta$. We will now see the comparison between cycloidal/trochoidal behavior and these forms of the Friedmann equations.

\section{Matter-dominated universe: cycloid}
\noindent
If we define a point on the circumference of a circle, and roll that circle along the plane in linear motion, the path traced by the individual positions of that point create the graph of a cycloid. 
\newline
\newline
A cycloid can be imagined as the graph of the continuous positions of a point $(0,0)$ on a circle of radius $r$, centered at $(0,r)$, as the circle rotates along the $x$ axis in the positive $x$ direction. If the circle rotates along the $x$ axis such that the center of the circle moves a distance of $l$ units from $(0,r)$ to $(l,r)$, then the circle rotates clockwise by an angle of $\frac{l}{r}=\theta$ radians to the vertical. As a result, the $x$ coordinate of the point $(0,0)$ increases by the distance traveled by the center, or $l=r\theta$ units, and decreases by the clockwise horizontal distance traveled from the circle's vertical diameter, or $r\sin(\theta)$. So, the $x$ coordinate can be represented as 
\begin{equation}
x=r(\theta-\sin(\theta)).
\label{cycloidxcoordinate}
\end{equation} 
Additionally, the $y$ coordinate of the point $(0,0)$ increases by the difference between the radius and clockwise vertical distance from the circle's horizontal diameter, or $r-r\cos(\theta)$. So, the $y$ coordinate can be represented as 
\begin{equation}
y=r(1-\cos(\theta)).
\label{cycloidycoordinate}
\end{equation}
Trochoids come as a variation of the cycloid theme, and help us reach the parametric equations for the beginning of the universe. Here, the impact of the cycloid and trochoid is the most important in determining the parametric equations.
\newline
\newline
We can replace the $y$ coordinate with $a$, the scale factor, the $x$ coordinate with $t$, time, and the constant $r$ with $C$ from Equation (\ref{matterdominated}). This alteration to Equations (\ref{cycloidxcoordinate}) and (\ref{cycloidycoordinate}) from the cycloid model, then we obtain the possible parametric equations for $a$ and $t$, in terms of variable $\theta$, as
\begin{equation}
t=C(\theta-\sin(\theta)),
\label{mattertimeequation}
\end{equation}
\begin{equation}
a=C(1-\cos(\theta)).
\label{matterscalefactorequation}
\end{equation}
Using these equations, we can produce the following equations for derivatives with respect to $\theta$
\[
\frac{da}{d\theta}=C\sin(\theta)),
\]
\[
\frac{dt}{d\theta}=C(1-\cos(\theta)).
\]
Breaking up the expression $(\dot{a})^2+1$ into the equivalence 
\[
(\dot{a})^2+1=\frac{\big(\frac{da}{d\theta}\big)^2}{\big(\frac{dt}{d\theta}\big)^2}+1
\]
and subsequently plugging in the above derivatives gives
\[
(\dot{a})^2+1=\frac{(C\sin(\theta))^2}{(C(1-\cos(\theta)))^2}+1=\frac{\sin^2(\theta)}{(1-\cos(\theta))^2}+1=\frac{(\sin^2(\theta))+(1-\cos(\theta))^2}{(1-\cos(\theta))^2},
\]
\[
(\dot{a})^2+1=\frac{(\sin^2(\theta))+(1-2\cos(\theta)+\cos^2(\theta))}{(1-\cos(\theta))^2}=\frac{\sin^2(\theta)+\cos^2(\theta)+1-2\cos(\theta)}{(1-\cos(\theta))^2}.
\]
\newline
Using the Pythagorean Identity $\sin^2(\theta)+\cos^2(\theta)=1$, we obtain
\[
(\dot{a})^2+1=\frac{1+1-2\cos(\theta)}{(1-\cos(\theta))^2}=\frac{2(1-\cos(\theta))}{(1-\cos(\theta))^2}=\frac{2}{1-\cos(\theta)}=\frac{2C}{C(1-\cos(\theta))}.
\]
Noting that the denominator of the last expression in the above equation equals Equation (\ref{matterscalefactorequation}), we derive \cite{LL, Lor}
\[
(\dot{a})^2+1=\frac{2C}{a}.
\]
We have just derived Equation (\ref{matterdominated}), which is the Friedmann equation for the non-relativistic, matter-dominated era of the universe's expansion, using the parametric equations of a cycloid. It turns out the cycloid is a perfect solution to the model given by the Friedmann equation for the non-relativistic, matter-dominated era. A note-worthy identity that can be realized from the equations for $a$ and $t$ is that the equations for $\frac{dt}{d\theta}$ and $a$ are identical, thus $a=\frac{dt}{d\theta}$.
\newline
\newline
A problem that arises with the cycloid solution is that there exists periodic points where $a=0$, at the minimum cusps in a cycloid graph. Thus, all of the matter and energy in the universe is being compressed into an infinitesimal point, indicative of infinite density. The Einstein-Cartan theory of gravity solves this problem. This theory extends general relativity by removing its symmetry constraint on the affine connection, which changes the conservation law for the orbital angular momentum into the conservation law for the orbital plus spin angular momentum \cite{ScKi,req}. The Einstein-Cartan theory thus admits the exchange between the orbital and spin components of angular momentum (spin–orbit interaction) in curved space-time, which is a result of the Dirac equation in relativistic quantum mechanics. At extremely high densities, torsion is a gravitationally repulsive interaction eliminating singularities \cite{avert,He,Kuc,Gas,Niko}. We will see this phenomenon applied in the next section, where a novel solution is presented.
\section{Torsion-dominated universe: trochoid}
\noindent
Alterations will be performed to Equation (\ref{radiationdominated}), since the torsion-dominated era at the beginning of the universe would have had no matter present, only pure energy and radiation. The torsion factor is inversely proportional to $a^4$, giving the Friedmann equation for this era \cite{Kuc,Gas,Niko,ApJ,SD} as
\begin{equation}
(\dot{a})^2+1=\frac{2A}{a^2}-\frac{(A^2-B^2)}{a^4}.
\label{torsiondominated}
\end{equation}
Notice how having $A\gtrapprox B$ results in the torsion factor making a significant impact on the function only for incredibly small values of $a$, which occurred in the tiny moments following the beginning of time. Here, instead of a singularity being formed as in the Big Bang Theory, a bounce is created at the minimum value of the universe's size, since the function never reaches the value of 0. This creates an alternative Big Bounce. The torsion factor approaches 0 rapidly after the Big Bounce, so the term $\frac{2A}{a^2}$ makes the only impact following this event. As a result, after the Big Bounce occurs, the torsion factor can be eliminated from the model without significant alterations in the results.
\newline
\newline
Now, we can alter the cycloid function to reflect this change, in that we can make the cycloid function not have $a$ equal to 0 at any time $t$. Instead of Equations (\ref{mattertimeequation}) and (\ref{matterscalefactorequation}) having the same coefficient for all expanded terms, we can make each term have a different coefficient. The new parametric equations are then
\begin{equation}
t=A\theta-B\sin(\theta),
\label{trochoidtime}
\end{equation}
\begin{equation}
a=A-B\cos(\theta)
\label{trochoidscalefactor}.
\end{equation}
Note that as long as positive constants $A>B$, the function for $a$ will always take a value greater than 0. Given the condition of $A\gtrapprox B$ from the above alteration of the Friedmann equation, $a$ will reach very close to 0, but never actually meet it. As a result, a singularity never forms with this model, creating the Big Bounce model discussed earlier. Note how the equivalence $a=\frac{dt}{d\theta}$ is preserved in these new alterations.
\newline
\newline
Here, we have formed the basic model of a trochoid, a closely related function to cycloids. This form of trochoids assumes only positive values, which is required for a real universe that never reaches singularities. A slight variation on the above trochoidal equation for $a$ by writing the equation in $a^2$ form yields the equation
\begin{equation}
a^2=A+B\cos(2\theta).
\label{torsionscalefactorsquared}
\end{equation}
It was hypothesized that Equation (\ref{torsionscalefactorsquared}) is a parametric model of the Friedmann equation from Equation (\ref{torsiondominated}), in that equation (\ref{torsionscalefactorsquared}) is a solution to the torsion-dominated era of the universe's expansion. This hypothesis yields 
\[
a^4=(A+B \cos(2\theta))^2,
\]
\[
\frac{da}{d\theta}=\frac{-2B\sin(\theta)\cos(\theta)}{\sqrt{A+B\cos(2\theta)}}.
\]
Squaring the above equation, and using the half-angle identities $\sin(\theta)=\sqrt{\frac{1-\cos(2\theta)}{2}}$ and $\cos(\theta)=\sqrt{\frac{1+\cos(2\theta)}{2}}$ yields
\[
\big(\frac{da}{d\theta}\big)^2=\frac{B^2(1-\cos^2(2\theta))}{A+B \cos(2\theta)}.
\]
Plugging these above results into
\[
(\dot{a})^2+1=\frac{\big(\frac{da}{d\theta}\big)^2}{\big(\frac{dt}{d\theta}\big)^2}+1=\frac{2A}{a^2}-\frac{(A^2-B^2)}{a^4}
\]
from Equation (\ref{torsiondominated}) yields

\[
\frac{\big(\frac{B^2(1-\cos^2(2\theta))}{A+B \cos(2\theta)}\big)}{\big(\frac{dt}{d\theta}\big)^2}+1=\frac{2A}{A+B \cos(2\theta)}-\frac{(A^2-B^2)}{(A+B \cos(2\theta))^2}=\frac{A^2+2AB\cos(2\theta)+B^2}{(A+B \cos(2\theta))^2},
\]
\[
\frac{\big(\frac{B^2(1-\cos^2(2\theta))}{A+B \cos(2\theta)}\big)}{\big(\frac{dt}{d\theta}\big)^2}=\frac{(A^2+2AB\cos(2\theta)+B^2)-((A+B \cos(2\theta))^2)}{(A+B \cos(2\theta))^2},
\]
\[
\frac{\big(\frac{B^2(1-\cos^2(2\theta))}{A+B \cos(2\theta)}\big)}{\big(\frac{dt}{d\theta}\big)^2}=\frac{(A^2+2AB\cos(2\theta)+B^2)-(A^2+2AB\cos(2\theta)+B^2\cos^2(2\theta))}{(A+B \cos(2\theta))^2}=\frac{B^2(1-\cos^2(2\theta))}{(A+B \cos(2\theta))^2},
\]
\[
\frac{1}{\big(\frac{dt}{d\theta}\big)^2}=\frac{1}{A+B \cos(2\theta)},
\]
\begin{equation}
\big(\frac{dt}{d\theta}\big)^2=A+B \cos(2\theta),
\label{blob8}
\end{equation}
\begin{equation}
t=\int{\big(\sqrt{A+B \cos(2\theta)}\big)}d\theta.
\label{blob9}
\end{equation}
The solution to this equation involves the usage of an elliptic integral of the second kind, denoted as
\begin{equation}
E(\theta,\psi)=\int_0^\theta{\sqrt{1-\psi^2\sin^2(\phi)}}d\phi.
\label{elliptic}
\end{equation}
\newline
The solution to Equation (\ref{blob9}) is
\begin{equation}
t=\sqrt{A+B}\cdot E(\theta,\psi),
\label{timeellipticintegral}
\end{equation}
\newline
where $\psi^2=\frac{2B}{A+B}$.
\newline
\newline
The hypothesis for $a(\theta)$ and the resulting equation for $t(\theta)$ will now be checked in the original Equation (\ref{torsiondominated}) to determine accuracy. We wish to show that the usage of Equation (\ref{torsionscalefactorsquared}) and Equation (\ref{timeellipticintegral}), when substituted into $(\dot{a})^2+1$, yield $\frac{2A}{a^2}-\frac{(A^2-B^2)}{a^4}$, representing the original Friedmann equation from Equation (\ref{torsiondominated}).
\newline
\newline
Plugging in the above equations for $\big(\frac{dt}{d\theta}\big)^2$ and $\big(\frac{da}{d\theta}\big)^2$ into the expanded form of $(\dot{a})^2+1$ yields 
\[
(\dot{a})^2+1=\frac{\big(\frac{da}{d\theta}\big)^2}{\big(\frac{dt}{d\theta}\big)^2}+1=\frac{\big(\frac{B^2(1-\cos^2(2\theta))}{A+B \cos(2\theta)}\big)}{\big(A+B \cos(2\theta)\big)}+1=\frac{B^2(1-\cos^2(2\theta))}{\big(A+B \cos(2\theta)\big)^2}+1=\frac{B^2(1-\cos^2(2\theta))+\big(A+B \cos(2\theta)\big)^2}{\big(A+B \cos(2\theta)\big)^2},
\]
\[
(\dot{a})^2+1=\frac{A^2+2AB\cos(2\theta)+B^2}{\big(A+B \cos(2\theta)\big)^2}=\frac{A^2+2AB\cos(2\theta)+A^2-A^2+B^2}{\big(A+B \cos(2\theta)\big)^2},
\]
\[
(\dot{a})^2+1=\frac{2A(A+B\cos(2\theta))}{\big(A+B \cos(2\theta)\big)^2}-\frac{(A^2-B^2)}{\big(A+B \cos(2\theta)\big)^2}=\frac{2A}{A+B \cos(2\theta)}-\frac{(A^2-B^2)}{\big(A+B \cos(2\theta)\big)^2}.
\]
\newline
Substituting Equation (\ref{torsionscalefactorsquared}) into the above equation yields
\[
(\dot{a})^2+1=\frac{2A}{a^2}-\frac{(A^2-B^2)}{a^4}.
\]
\newline
We have derived Equation (\ref{torsiondominated}), the Friedmann equation representing the expansion of the universe when it was infinitesimally small and infinitesimally young, from parametric equations for $a$ and $t$ given by Equations (\ref{torsionscalefactorsquared}) and (\ref{timeellipticintegral}). So, we haven proven that in the beginning conditions of the positively curved universe, satisfied by the equation $(\dot{a})^2+1=\frac{2A}{a^2}-\frac{(A^2-B^2)}{a^4}$, scale factor $a$ and time $t$ can be represented by parametric equations in terms of $\theta$, as $a^2=A+B \cos(2\theta)$ and $t=\sqrt{A+B}\cdot E(\theta,\psi)$, where $\psi^2=\frac{2B}{A+B}$. These parametric equations constitute a novel solution to the Friedmann equation for the torsion-dominated early universe.

\section{Radiation-dominated universe}
\noindent
Within a small amount of time into the initial expansion of the universe, the scale factor $a$ surges massively upward, causing the torsion factor to rapidly approach 0. The torsion-dominated Friedmann equation from Equation (\ref{torsiondominated}) gives way to the original relativistic, radiation-dominated universe Friedmann equation from Equation (\ref{radiationdominated}), when the torsion factor $\frac{(A^2-B^2)}{a^4}$ equals 0. This occurs when the numerator equals 0, when positive constants $A$ and $B$ are equal to each other. Then, Equation (\ref{torsionscalefactorsquared}) turns into
\[
a^2=A(1+ \cos(2\theta)).
\]
With the substitution of the cosine-function double-angle-identity, $\cos(2\theta)=2\cos^2(\theta)-1$, we obtain
\begin{equation}
a^2=2A\cos^2(\theta),
\label{earlyscalefactorsquared}
\end{equation}
\begin{equation}
a=\sqrt{2A}\cos(\theta),
\label{earlystagesa}
\end{equation}
\begin{equation}
\frac{da}{d\theta}=-\sqrt{2A}\sin(\theta),
\label{earlystagesda/dtheta}
\end{equation}
\begin{equation}
\big(\frac{da}{d\theta}\big)^2=2A\sin^2(\theta).
\label{earlystagesda/dthetasquared}
\end{equation}
Substituting the result $A=B$ into Equation (\ref{blob9}) yields
\[
t=\int\big({\sqrt{A(1+\cos(2\theta))}\big)d\theta}.
\]
Again, substituting the cosine-function double-angle-identity, $\cos(2\theta)=2\cos^2(\theta)-1$, yields
\begin{equation}
t=\int\big({\sqrt{2A\cos^2(\theta)}\big)d\theta}=\int\big({\sqrt{2A}\cos(\theta)\big)d\theta}=\sqrt{2A}\sin(\theta),
\label{earlystagest}
\end{equation}
\begin{equation}
\frac{dt}{d\theta}=\sqrt{2A}\cos(\theta),
\label{earlystagesdt/dtheta}
\end{equation}
\begin{equation}
\big(\frac{dt}{d\theta}\big)^2=2A\cos^2(\theta).
\label{earlystagesdt/dthetasquared}
\end{equation}
Plugging in the above equations for $\big(\frac{dt}{d\theta}\big)^2$ and $\big(\frac{da}{d\theta}\big)^2$ into the expanded form of $(\dot{a})^2+1$ yields 
\[
(\dot{a})^2+1=\frac{\big(\frac{da}{d\theta}\big)^2}{\big(\frac{dt}{d\theta}\big)^2}+1=
\frac{2A\sin^2(\theta)}{2A\cos^2(\theta)}+1=\tan^2(\theta)+1.
\]
Using the Pythagorean Identity $\tan^2(\theta)+1=\sec^2(\theta)$ then yields
\[
(\dot{a})^2+1=\sec^2(\theta)=\frac{1}{\cos^2(\theta)}=\frac{2A}{2A\cos^2(\theta)}.
\]
Substituting Equation (\ref{earlyscalefactorsquared}) then yields
\[
(\dot{a})^2+1=\frac{2A}{a^2}.
\]
We have derived Equation (\ref{radiationdominated}), the Friedmann equation representing the expansion of the relativistic, radiation-dominated universe, during the time period following significant impact of the torsion factor. Note how the equivalence $a=\frac{dt}{d\theta}$ is still preserved in this new set of solutions for the radiation-dominated era.

\section{Dark energy universe}
\noindent
One further analysis can be put forth for completion of all possible analyses of the Friedmann equations, that being the current, dark-energy-dominated universe. This exists in the matter-dominated time period, so Equation (\ref{matterdominated}) will be altered to add in the dark energy factor. This factor is directly proportional to $a^2$, giving the equation
\begin{equation}
(\dot{a})^2+1=\frac{2C}{a}+Da^2
\label{darkenergy}
\end{equation}
Further into the evolution of the universe, as in our current time period and the future, the scale factor is an extremely large number. This means that $\frac{2C}{a}$ approaches 0, and makes little contribution. The function converges to 
\[
(\dot{a})^2=Da^2,
\]
\[
\dot{a}=\sqrt{D}a,
\]
\[
a=e^{t\sqrt{D}}.
\]
The equation converges to the exponential function of increasing scale factor that is reflected in our current cosmological observations, and the universe increases in size indefinitely.
\newline
\newline
An important analysis resulting from this model is that the value of the constant $D$ must be greater than some threshold value, which depends on $C$, in order for the universe to start acceleration.

\section{Summary}
\noindent
We have developed novel parametric solutions for scale factor and time for the torsion-dominated era of the universe, near the Big Bounce, given positive closed curvature. The Friedmann equation used for this specific analysis is
\[
(\dot{a})^2+1=\frac{2A}{a^2}-\frac{(A^2-B^2)}{a^4},
\]
\newline
with parametric solutions as 
\[
a^2=A+B \cos(2\theta),
\]
\[
t=\sqrt{A+B}*E(\theta,\psi),
\]
where $\psi^2=\frac{2B}{A+B}$.
\newline
\newline
This solution can be used with the condition $A=B$ to solve for the parametric equations of the relativistic, radiation-dominated era of the universe, at which point the torsion factor added to the Friedmann equation zeroes out. Moreover, the inspiration for such parametric solutions comes from the basis of cycloidal/trochoidal behavior of a positively curved universe, given the parametric equations of the matter-dominated, non-relativistic era of the universe being equal to those of a cycloidal function.
\newline
\newline
Important conclusions that can be made are that there may have been a "bounce" period near the beginning of the universe, at which point the universe had never formed a singularity, but rather "bounced" in a state extremely close to infinite density. This model suggests that the universe never had a Big Bang, or a definite beginning of time for that matter, and the beginning of time cannot be traced back to the most recent absolute minimum size of the universe, due to the "bounce" period described here. Rather, the Big Bounce acted as an inflection point between two successive stages of evolution. Also, the models demonstrate that the coefficients of the Friedmann equation forms become irrelevant at extremely large values of scale factor and expansion rate in a universe governed by dark energy, making the fate of such closed universes to inevitably expand forever. An important result is that the closed universe model follows cycloidal/trochoidal behavior, as is represented with the scenarios where parametric equations involving correlated trigonometric functions were found as solutions to the Friedmann equations. 
\newline
\newline
An important finding that comes up in these analyses is that if the Big Bounce had indeed occurred, then the previous evolutionary stage of the universe, prior to the most recent Big Bounce, must not have been able to enter the dark energy phase of evolution and increase in size exponentially and indefinitely, thus reentering bounce periods. Since we can rule out the possibility that dark energy can never overtake ordinary matter's gravitational pull because of scientific evidence against it, the model suggests that the influences of dark energy must increase for each successive developmental stage of the universe. This means that the effects of dark energy increase over successive evolutionary courses of the universe, each separated by a bounce period. In this present stage, the universe must have reached, for the first time and possibly only time, an evolutionary stage preceded by a bounce period but overtaken by dark energy and expanding indefinitely forever.
\newline
\newline
These analyses remain critical for further research in cosmological evolution and the closed universe model. Overall, the elegance of the Friedmann equations in this setting are highly demonstrated by these analyses.
\section*{Acknowledgments}
\noindent
I thank Professor Nikodem Poplawski of the University of New Haven for supervising and guiding me with my research project, and for his continued support in this work. I also thank Professor Joseph Kolibal for helping me initiate my research and providing me with helpful advice throughout my project.

\end{document}